# Commparison of the Hanbury Brown-Twiss effect for bosons and fermions

T. Jeltes[1], J. M. McNamara[1], W. Hogervorst[1], W. Vassen[1], V. Krachmalnicoff[2], M. Schellekens[2], A. Perrin[2], H. Chang[2], D. Boiron[2], A. Aspect[2], C. I. Westbrook[2]

[1]*Laser Centre Vrije Universiteit, De Boelelaan 1081, 1081 HV Amsterdam, The Netherlands, and* [2]*Laboratoire Charles Fabry de l'Institut d'Optique, CNRS, Univ Paris-sud, Campus Polytechnique RD 128, 91127 Palaiseau Cedex, France.*

**Fifty years ago, Hanbury Brown and Twiss (HBT) discovered photon bunching in light emitted by a chaotic source[1], highlighting the importance of two-photon correlations[2] and stimulating the development of modern quantum optics[3]. The quantum interpretation of bunching relies upon the constructive interference between amplitudes involving two indistinguishable photons, and its additive character is intimately linked to the Bose nature of photons. Advances in atom cooling and detection have led to the observation and full characterisation of the atomic analogue of the HBT effect with bosonic atoms[4,5,6]. By contrast, fermions should reveal an antibunching effect, *i.e.*, a tendency to avoid each other. Antibunching of fermions is associated with destructive two-particle interference and is related to the Pauli principle forbidding more than one identical fermion to occupy the same quantum state. Here we report an experimental comparison of the fermion and the boson HBT effects realised in the same apparatus with two different isotopes of helium, $^{3}$He (a fermion) and $^{4}$He (a boson). Ordinary attractive or repulsive interactions between atoms are negligible, and the contrasting bunching and antibunching behaviours can be fully attributed to the different quantum statistics. Our result shows how atom-atom correlation measurements can be used not only for revealing details in the spatial density[7,8] or momentum correlations[9] in an atomic ensemble, but also to directly observe phase**

**effects linked to the quantum statistics in a many body system. It may thus find**
**applications to study more exotic situations[10].**

Two-particle correlation analysis is an increasingly important method for studying
complex quantum phases of ultracold atoms[7,8,9,10,11,12,13]. It goes back to the discovery
by Hanbury Brown and Twiss[1], that photons emitted by a chaotic (incoherent) light
source tend to be bunched: the joint detection probability is enhanced, compared to that
of statistically independent particles, when the two detectors are close together.
Although the effect is easily understood in the context of classical wave optics[14], it took
some time to find a clear quantum interpretation[3,15]. The explanation relies upon
interference between the quantum amplitude for two particles, emitted from two source
points $S_1$ and $S_2$, to be detected at two detection points $D_1$ and $D_2$ (see fig. 1). For
bosons, the two amplitudes $\langle D_1 | S_1 \rangle \langle D_2 | S_2 \rangle$ and $\langle D_1 | S_2 \rangle \langle D_2 | S_1 \rangle$ must be added,
which yields a factor of 2 excess in the joint detection probability, if the two amplitudes
have the same phase. The sum over all pairs $(S_1, S_2)$ of source points washes out the
interference, unless the distance between the detectors is small enough that the phase
difference between the amplitudes is less than one radian, or equivalently if the two
detectors are separated by a distance less than the coherence length. Study of the joint
detection rates *vs.* detector separation along the *i*-direction then reveals a bump whose
width $l_i$ is the coherence length along that axis[1,5,16,17,18,19]. For a source size $s_i$ along *i*
(standard half width at $e^{-1/2}$ of a Gaussian density profile), one has a half width at $1/e$ of
$l_i = \mathrm{h}t / 2\pi m s_i$, where *m* is the mass of the particle, *t* the time of flight from the source to
the detector, and *h* Planck's constant. This formula is the analogue of the formula
$l_i = L\lambda / 2\pi s_i$ for photons if one identifies $\lambda = h / mv$ with the de Broglie wavelength for
particles travelling at velocity $v = L / t$ from the source to the detector.

For indistinguishable fermions, the two-body wave function is antisymmetric, and
the two amplitudes must be subtracted, yielding a null probability for joint detection in
the same coherence volume. In the language of particles, it means that two fermions
cannot have momenta and positions belonging to the same elementary cell of phase
space. As a result, for fermions the joint detection rate *vs.* detector separation is
expected to exhibit a dip around the null separation. Such a dip for a fermion ensemble must not be confused with the antibunching dip that one can observe with a single particle (boson or fermion) quantum state, *e.g.,* resonance fluorescence photons emitted by an individual quantum emitter[20]. In contrast to the HBT effect for bosons, the fermion analogue cannot be interpreted by any classical model, either wave or particle, and extensive efforts have been directed toward an experimental demonstration. Experiments have been performed with electrons in solids[21,22] and in a free beam[23], and with a beam of neutrons[24], but none has allowed a detailed study and a comparison of the pure fermionic and bosonic HBT effects for an ideal gas. A recent experiment using fermions in an optical lattice[25] however, does permit such a study and is closely related to our work.

Here, we present an experiment in which we study the fermionic HBT effect for a sample of polarised, metastable $^3$He atoms ($^3$He*), and we compare it to the bosonic HBT effect for a sample of polarised, but not Bose condensed, metastable $^4$He atoms ($^4$He*) produced in the same apparatus at the same temperature. We have combined the position and time resolved detector previously used[5,26] for $^4$He*, with an apparatus with which ultracold samples of $^3$He* or $^4$He* have recently been produced[27]. Fermions or bosons at thermal equilibrium in a magnetic trap are released onto the detector which counts individual atoms (see Fig. 1) with an efficiency of approximately 10%. The detector allows us to construct the normalised correlation function $g^{(2)}(\Delta\mathbf{r})$, *i.e.* the probability of joint detection at two points separated by $\Delta\mathbf{r}$, divided by the product of the single detection probabilities at each point. Statistically independent detection events result in a value of 1 for $g^{(2)}(\Delta\mathbf{r})$. A value larger than 1 indicates bunching, while a value less than 1 is evidence of antibunching.

We produce gases of pure $^3$He* or pure $^4$He* by a combination of evaporative and sympathetic cooling in an anisotropic magnetic trap (see Methods). Both isotopes are in pure magnetic substates, with nearly identical magnetic moments and therefore nearly identical trapping potentials, so that trapped non degenerate and non interacting samples have the same size at the same temperature. The temperatures of the samples yielding the results of Fig. 2, as measured by the spectrum of flight times to the detector, are 0.53 ± 0.03 µK and 0.52±0.05 µK for $^3$He* and $^4$He* respectively. The uncertainties correspond to the standard deviation of each ensemble. In a single realisation, we typically produce $7 \times 10^4$ atoms of both $^3$He* and $^4$He*. The atom number permits an estimate of the Fermi and BEC temperatures of approximately 0.9 µK and 0.4 µK respectively. Consequently, Fermi pressure in the trapped $^3$He* sample has a negligible (3%) effect on the trap size and repulsive interactions in the $^4$He* sample have a similarly small effect. The trapped samples are therefore approximately Gaussian ellipsoids elongated along the *x*-axis with an *rms* size of about 110×12×12 µm$^3$. To release the atoms, we turn off the current in the trapping coils and atoms fall under the influence of gravity. The detector, placed 63 cm below the trap centre (see fig. 1), then records the *x-y* position and arrival time of each detected atom.

The normalised correlation functions g$^{(2)}$ (0,0,Δz) along the *z* (vertical) axis, for $^3$He* and $^4$He* gases at the same temperature, are shown in Fig 2. Each correlation function is obtained by analysing the data from about 1000 separate clouds for each isotope (see Methods). Results analogous to those of Fig. 2 are obtained for correlation functions along the *y*-axis, but the resolution of the detector in the *x-y* plane (about 500 µm halfwidth at 1/e for pair separation) broadens the signals. Along the *x*-axis (the long axis of the trapped clouds), the expected widths of the HBT structures are one order of magnitude smaller than the resolution of the detector and are therefore not resolved.

Figure 2 shows clearly the contrasting behaviours of bosons and fermions. In both cases one observes a clear departure from statistical independence at small separation. Around zero separation, the fermion signal is lower than unity (antibunching) while the boson signal is higher (bunching). Since the sizes of the $^3$He* and $^4$He* clouds at the same temperature are the same, as are the times of flight (pure free fall), the ratio of the correlation lengths is expected to be equal to the inverse of the mass ratio, 4/3. The observed ratio of the correlation lengths along the *z* axis in the data shown is 1.3 ±0.2. The individual correlation lengths are also in good agreement with the formula $l_z=ht/2\pi m s_z$. Due to the finite resolution, the contrast in the signal, which should ideally go to 0 or 2 is reduced by a factor of order ten. The amount of contrast reduction is slightly different for bosons and fermions and the ratio should be about 1.5. The measured ratio is 2.4 ± 0.2. This discrepancy has several possible explanations. First, the magnetic field switch-off is not sudden (time scale ~1 ms) and this could affect bosons and fermions differently. Second, systematic errors may be present in our estimate of the resolution function. The resolution however, does not affect the widths of the observed correlation functions along $z$, and thus we place the strongest emphasis on this ratio as a test of our understanding of boson and fermion correlations in an ideal gas. More information on uncertainties, systematic errors as well as a more complete summary of the data are given in the supplementary material.

Improved detector resolution would allow a more detailed study of the correlation function, and is thus highly desirable. One can circumvent the effect of the resolution using a diverging atom lens to demagnify the source[4]. According to the formula $l = ht/2\pi ms$, a smaller effective source size gives a larger correlation length. We have tried such a scheme by creating an atomic lens with a blue detuned, vertically propagating, laser beam, forcing the atoms away from its axis (see Methods). The laser waist was not large compared to the cloud size and therefore our "lens" suffered from strong aberrations, but a crude estimate of the demagnification, neglecting aberrations, gives about 2 in the $x$-$y$ plane. Figure 3 shows a comparison of $g^{(2)}(\Delta z)$ for fermions with and without the defocusing lens. We clearly see a greater antibunching depth, consistent with larger correlation lengths in the $x$-$y$ plane (we have checked that $l_y$ is indeed increased) and therefore yielding a smaller reduction of the contrast when convolved with the detector resolution function. As expected, the correlation length in the $z$-direction is unaffected by the lens in the $x$-$y$ plane. Although our atomic lens was far from ideal, the experiment shows that it is possible to modify the HBT signal by optical means.

To conclude, we emphasise that we have used samples of neutral atoms at a moderate density in which interactions do not play any significant role. Care was taken to manipulate bosons and fermions in conditions as similar as possible. Thus the observed differences can be understood as a purely quantum effect associated with the exchange symmetries of wave functions of indistinguishable particles.

The possibility of having access to the sign of phase factors in a many body wave function opens fascinating perspectives for the investigation of intriguing analogues of condensed matter systems, which can now be realised with cold atoms. For instance, one could compare the many body state of cold fermions and that of "fermionised" bosons in a 1D sample[28,29]. Our successful manipulation of the HBT signal by interaction with a laser suggests that other lens configurations could allow measurements in position space (by forming an image of the cloud at the detector) or in any combination of momentum and spatial coordinates.

**Methods**

**Experimental sequence.** Clouds of cold $^4$He* are produced by evaporative cooling of a pure $^4$He* sample, loaded into a Ioffe-Pritchard magnetic trap[30]. The trapped state is $2^3S_1$, $m_J = 1$ and the trap frequency values are 47Hz and 440Hz, for axial and radial confinement respectively. The bias field is 0.75 G corresponding to a frequency of 2.1 MHz for a transition between the $m_J = 1$ and $m_J = 0$ states at the bottom of the trap. After evaporative cooling, we keep an RF knife on at constant frequency for 500 ms, then wait for 100 ms before switching off the trap. In contrast to the experiments of Ref. 5, atoms are released in a magnetic field sensitive state.

To prepare $^3$He* clouds, we simultaneously load $^3$He* and $^4$He* atoms in the magnetic trap[27]. The trapping state for $^3$He* is $2^3S_1$, F=3/2, $m_F$=3/2, and axial and radial trap frequencies are 54 Hz and 506 Hz – the difference compared to $^4$He* is only due to the mass. The two gases are in thermal equilibrium in the trap, so that $^3$He* is sympathetically cooled with $^4$He* during the evaporative cooling stage. Once the desired temperature is reached, we selectively eliminate $^4$He* atoms from the trap using the RF knife. The gyromagnetic ratios for $^4$He* and $^3$He* are 2 and 4/3 respectively, so that the resonant frequency of the m=1 to m=0 transition for $^4$He* is 3/2 times larger than the m=3/2 to m=1/2 transition for $^3$He*. An RF ramp from 3 MHz to 1.9 MHz expels all the $^4$He* atoms from the trap without affecting $^3$He*. We then use the same trap switch-off procedure to release the $^3$He* atoms onto the detector, also in a magnetic field sensitive state. We can apply magnetic field gradients to check the degree of spin polarisation of either species.

**Correlation Function.** The detailed procedure leading to this correlation is given in Ref. 5. Briefly, we convert arrival times to $z$ positions and then use the 3-dimensional positions of each atom to construct a histogram of pair separations Δ**r** in a particular cloud. We then sum the pair distribution histograms for 1000 successive runs at the same temperature. For separations much larger than the correlation length, this histogram reflects the Gaussian spatial distribution of the cloud. To remove this large scale shape and obtain the normalised correlation function we divide the histogram by the autoconvolution of the sum of the 1000 single particle distributions.

**Atom lens experiment.** A 300 mW beam with an elliptical waist of approximately $100\times150$ μm$^2$ propagates vertically through the trap. The laser frequency is detuned by 300 GHz from the $2^3S_1$ to $2^3P_2$ transition. After turning off the magnetic trap, and waiting 500 μs for magnetic transients to die away, the defocusing laser is turned on for 500 μs.

**Supplementary Information** accompanies the paper on **www.nature.com/nature**.

Acknowledgements: This work was supported by the access programme of LASERLAB EUROPE. The LCVU group in Amsterdam is supported by the "Cold Atoms" program of the Dutch Foundation for Fundamental Research on Matter (FOM) and by the Space Research Organization Netherlands (SRON). The Atom Optics group of LCFIO is member of the IFRAF instute and of the Fédération LUMAT of the CNRS, and is supported by the french ANR and by the SCALA programme of the European Union.

Competing Interests statement: The authors declare that they have no competing financial interests.

Correspondence and requests for materials should be addressed to christoph.westbrook@institutoptique.fr or w.vassen@few.vu.nl

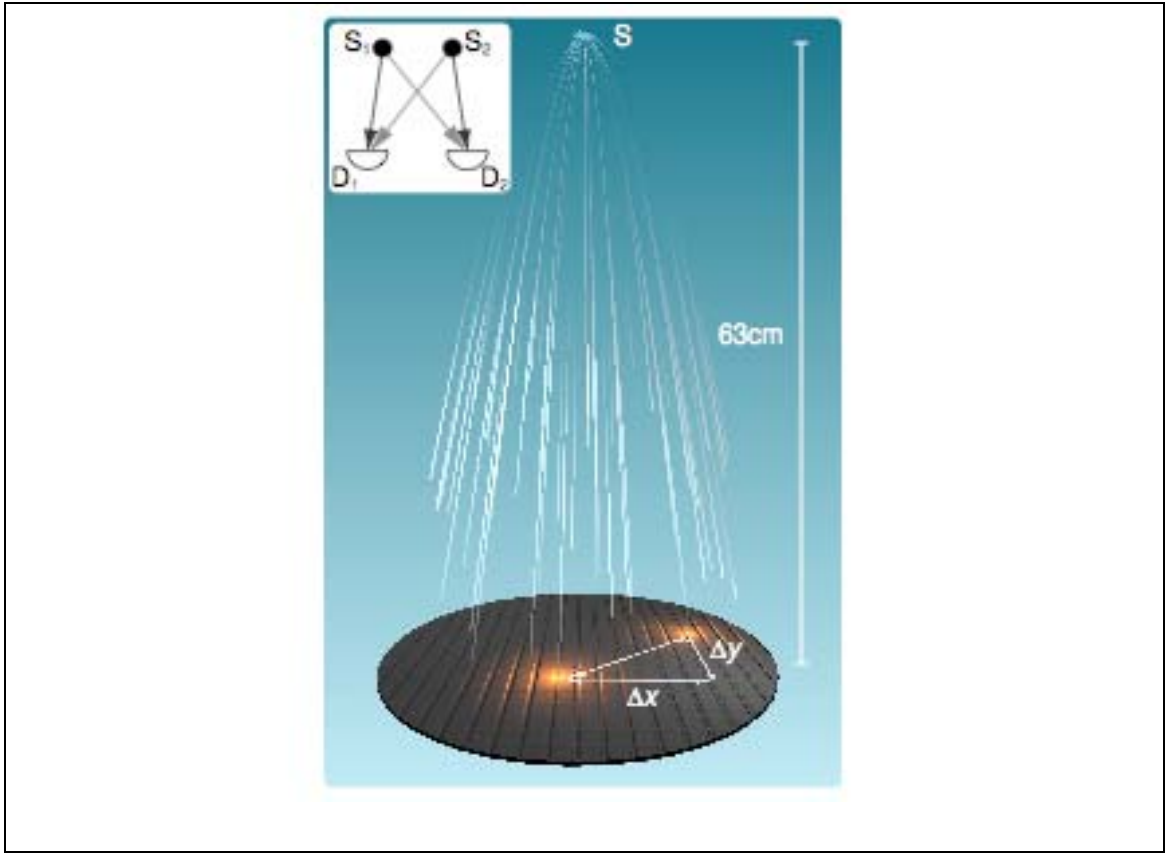

Caption for figure 1: The experimental setup. A cold cloud of metastable helium atoms is released at the switch-off of a magnetic trap. The cloud expands and falls under the effect of gravity onto a time resolved and position sensitive detector (micro-channel plate and delay-line anode), that detects single atoms. The inset shows conceptually the two 2-particle amplitudes (in black or grey) that interfere to give bunching or antibunching: $S_1$ and $S_2$ refer to the initial positions of two identical atoms jointly detected at $D_1$ and $D_2$.

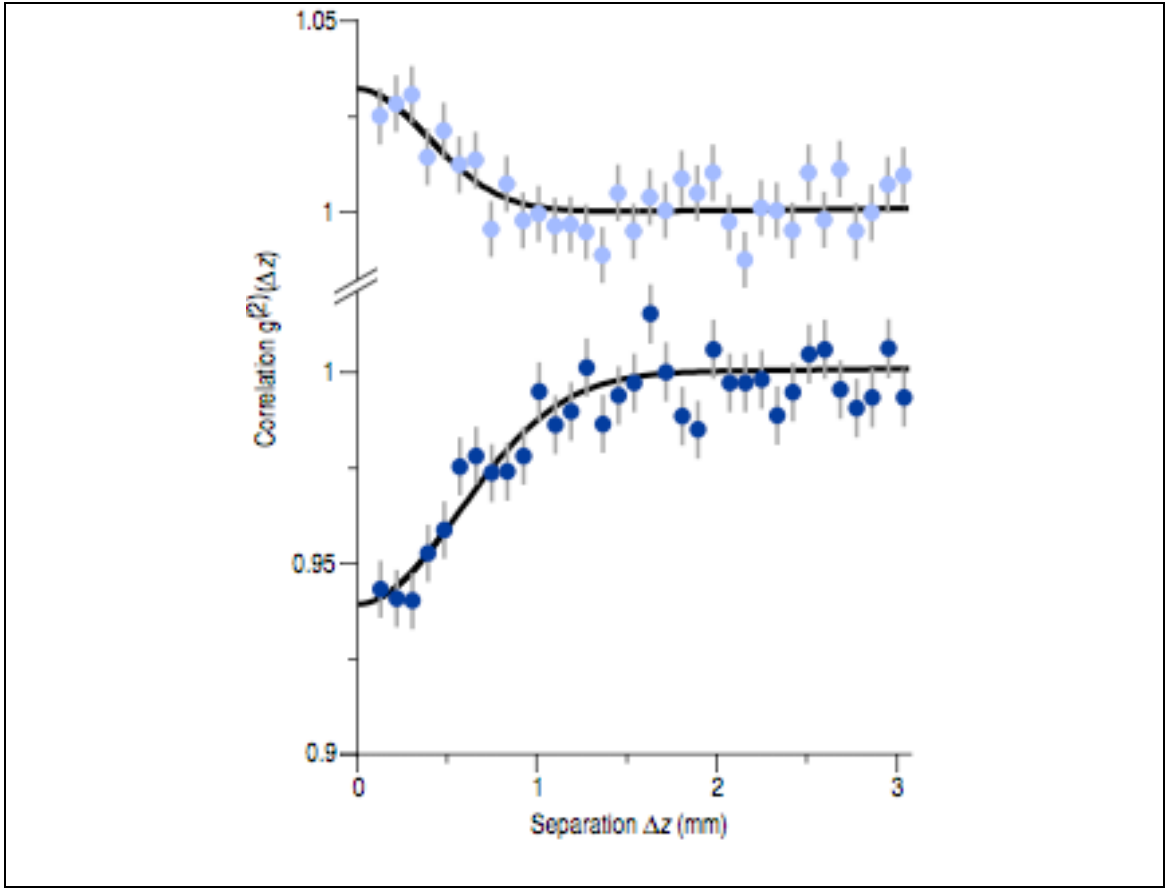

Caption for figure 2: Normalised correlation functions for $^4$He* (bosons) in the upper graph, and $^3$He* (fermions) in the lower graph. Both functions are measured at the same cloud temperature (0.5 µK), and with identical trap parameters. Error bars correspond to the root of the number of pairs in each bin. The line is a fit to a Gaussian function. The bosons show a bunching effect; the fermions anti-bunching. The correlation length for $^3$He* is expected to be 33% larger than that for $^4$He* due to the smaller mass. We find 1/e values for the correlation lengths of 0.75±0.07 mm and 0.56±0.08 mm for fermions and bosons respectively.

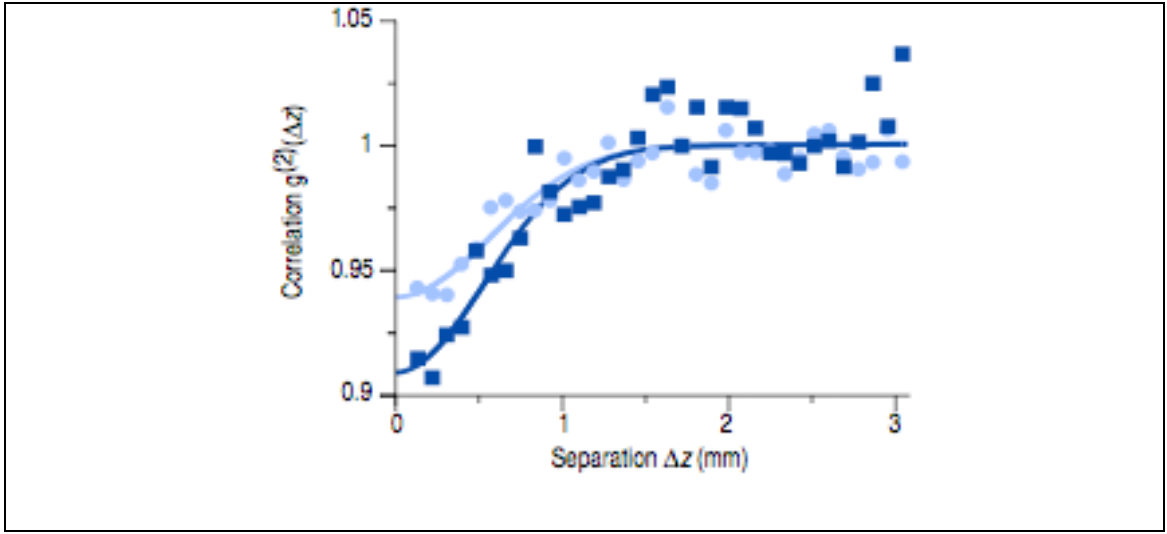

Caption for figure 3: Effect of demagnifying the source size. We show normalised correlation functions along the *z* (vertical) axis for $^3$He*, with (dark blue squares) and without (light blue circles) a diverging atomic lens in the *x-y* plane. The dip is deeper with the lens because of the increase of the correlation lengths in the *x-y* plane leading to less reduction of contrast when convolved by the resolution function in that plane.

Supplementary material for "Hanbury Brown Twiss effect for bosons versus fermions"

*1. Unnormalised pair histogram*

In order to give the reader an idea of the "raw" data, we show in Supplementary figure 1 some unnormalised pair histograms. The data correspond to the normalised plots shown in Fig. 2 in the main text. In addition to the bunching and antibunching feature for separations below 1 mm, the histogram also shows a broad structure which is due to the approximately Gaussian shape of the cloud. The broad structure is eliminated by the normalisation procedure described in Ref. 5 of the main text and summarised in Methods.

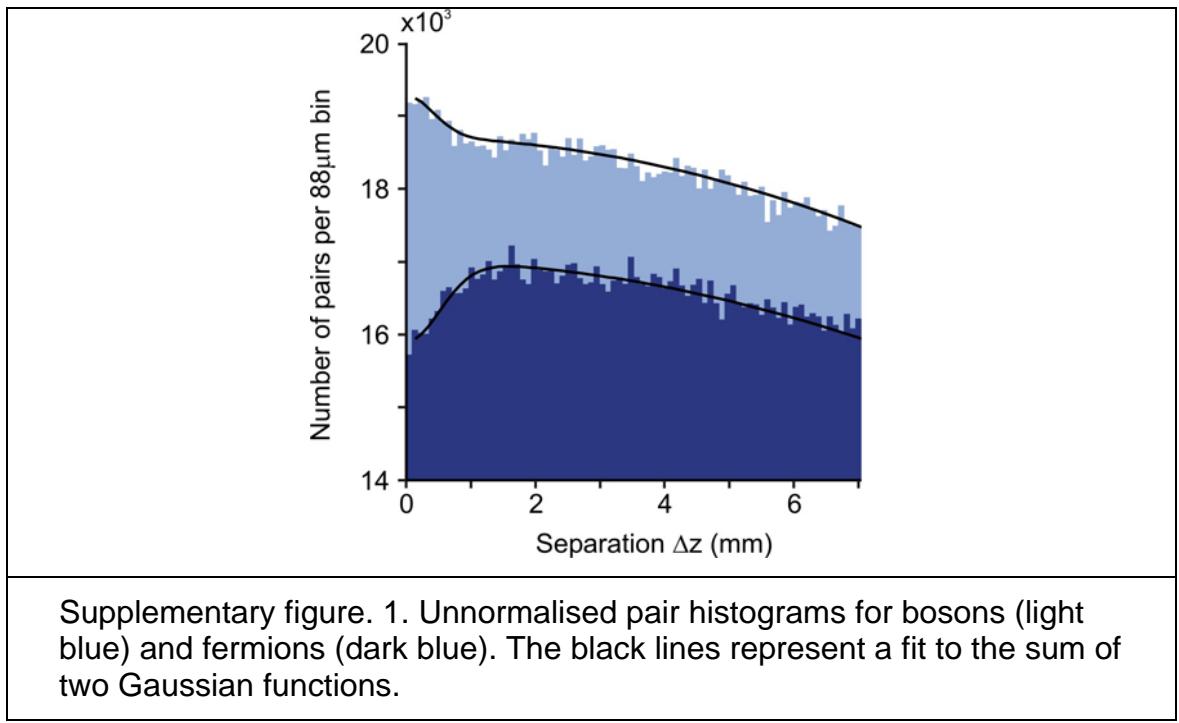

Supplementary figure. 1. Unnormalised pair histograms for bosons (light blue) and fermions (dark blue). The black lines represent a fit to the sum of two Gaussian functions.

## 2. Fit results

If one neglects finite resolution effects, the normalised correlation function should be well described by a Gaussian function:

$$g^{(2)}(\Delta x, \Delta y, \Delta z) = 1 \pm \exp\left(-\left[\left(\frac{\Delta x}{l_x}\right)^2 + \left(\frac{\Delta y}{l_y}\right)^2 + \left(\frac{\Delta z}{l_z}\right)^2\right]\right), \tag{1}$$

where the + sign refers to bosons and the – sign to fermions. We denote the correlation lengths in the 3 different spatial directions $i$, by $l_i$. In practice this function must be convolved with the resolution function of the detector. The resolution function is determined by the method discussed in Ref. 26. The resolution along the $z$ direction is approximately 3 nm and is neglected. The convolution in the $x$-$y$ plane is described in Ref. 19 for the case of a Gaussian resolution function. Careful measurements have revealed that the wings of the resolution function are broader than those of a Gaussian and we thus use an empirically determined analytical function to approximate the pair resolution function. Its $1/e$ halfwidth is about 500 µm. Since the correlation length in the $x$ direction is more than an order of magnitude smaller than the resolution, we set $l_x = 0$. The convolution also affects the height of the signal so that $g^{(2)}(0,0,0) = 1 \pm \eta$. The parameter $\eta$ is referred to as the contrast. The fit parameters are thus $l_y$, $l_z$ and $\eta$.

Data were taken for fermions ($^3$He*) at 0.5 µK, 1.0 µK and 1.4 µK. The corresponding fit results for $l_z$ and $\eta$ are plotted in Supplementary Figure 2. In addition, we have data for two other situations, one using $^4$He* at 0.5 µK, and another using $^3$He* at 0.5 µK and a diverging lens. All 5 runs are summarised in table 1. In the graphs, we have plotted the formula $l_z = ht/2\pi m s_z$, extracting the size $s_z$ from the measured temperature, trap oscillation frequency and assuming the cloud is an ideal gas. For the contrast $\eta$, we plot the expected variation based on the measured resolution function.

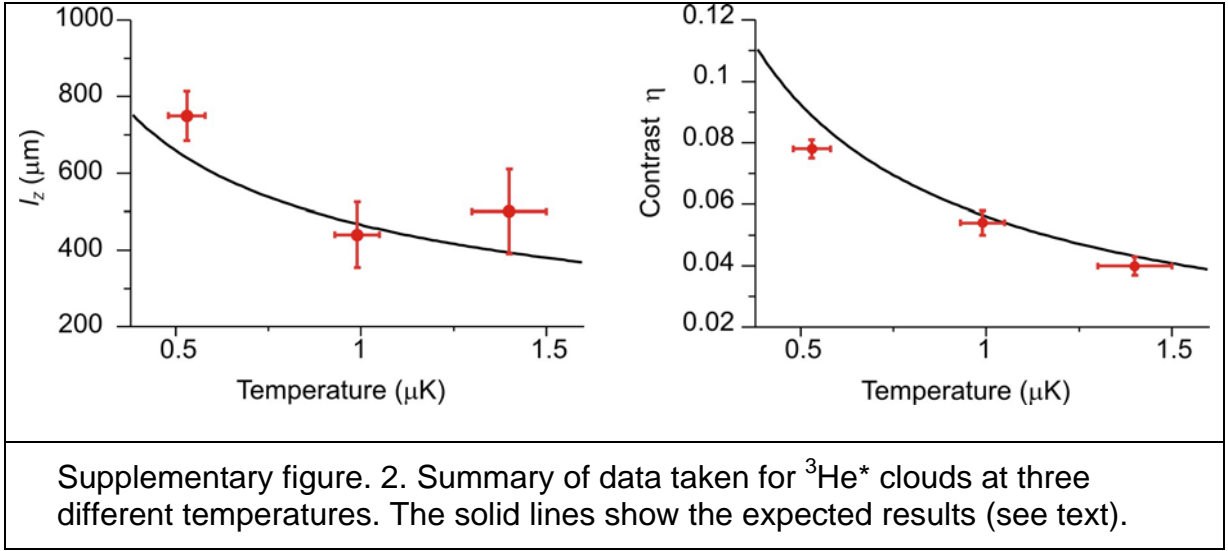

Supplementary figure. 2. Summary of data taken for $^3$He* clouds at three different temperatures. The solid lines show the expected results (see text).

| Run | $l_z$ (μm) | $l_y$ (μm) | η |
|---|---|---|---|
| $^3$He*, 0.5 μK | 750 ± 70 | 570 ± 50 | 0.078 ± 0.003 |
| $^3$He*, 1 μK | 440 ± 90 | 360 ± 90 | 0.054 ± 0.004 |
| $^3$He*, 1.4 μK | 500 ± 110 | 0 * | 0.040 ± 0.003 |
| $^4$He*, 0.5 μK | 560 ± 80 | 570 ± 100 | 0.033 ± 0.003 |
| $^3$He*, 0.5 μK, with lens | 750 ± 80 | 810 ± 40 | 0.108 ± 0.003 |

*In this run, the fitted width of the correlation function along $y$ is actually smaller than the resolution. Thus no reasonable value can be extracted for $l_y$.

Supplementary table 1. Summary of fit results for all data sets

Generally the data are in good agreement with the predictions of the ideal gas model. In the run with the lens, we have made no quantitative comparison with a calculation because it would involve taking into account the severe aberrations of the lens. Qualitatively however, we see that, as expected, the correlation length along $z$ is unchanged while that along $y$, as well as the contrast, are increased. The fitted values of η do not correspond to those one would deduce from the data in Figs. 2 and 3 in the main text. This is because, as in Ref. 5, we computed the correlation function along the z axis over an area slightly larger than the width of the resolution function. This procedure improves the signal to noise ratio and preserves the form and the width of the correlation function but slightly modifies its height.

We observe three small anomalies: first, the contrast η for both bosons and fermions at 0.5 µK is below the prediction and the ratio, after correction for the resolution, is 2.4 ± 0.2 instead of the expected value 1.5. Second, the correlation length $l_y$ for fermions at 0.5 µK seems quite low resulting in a ratio of fermions to bosons of 1.0 ± 0.2 instead of 1.3. Third, the width of the antibunching dip in the normalised pair separation histogram along y for fermions at 1.4 µK is *smaller* than the width of the measured resolution function, meaning that the fitted value of $l_y$ is consistent with zero.

A systematic error may be present in the estimation of the detector resolution. After moving the detector from Orsay to Amsterdam, we noticed that the detector resolution differed by up to 30% from day to day. A systematic error in the resolution has approximately the same relative effect on the value of η. It would also have an effect on the value of $l_y$. Uncontrolled variations in the resolution may thus account for the above anomalies. The correlation length in the vertical direction $l_z$ however, should not be affected by an imprecise knowledge of the resolution in the x-y plane. The good agreement we find with our expectations along this axis is the strongest argument that the correlations we observe are consistent with the ideal gas model.

A second possible source of systematic error is related to the switch-off of the magnetic trap. Eddy currents cause a typical time scale of 1 ms in this turn-off[30]. Since, unlike in Ref. 5, the released atoms are in a magnetic field sensitive state, partially adiabatic effects or focussing by residual curvatures could affect our measurement of the temperature or of the effective source size viewed from the detector. We have no independent estimate of the magnitude of these effects and can simply conclude that the reasonable agreement with our model means that these effects are not very large.